%% file: paper.tex
\documentclass[10pt,twocolumn,journal]{IEEEtran}

\usepackage{amsmath, mathrsfs, mathtools}
\usepackage{amsfonts}
\usepackage{amssymb}
\usepackage{color}
\usepackage{multirow}
\usepackage{stmaryrd}
\usepackage{mathabx}

\usepackage[
a4paper,
left=1.65cm,
right=1.65cm,
top=1.65cm,
bottom=1.65cm]
{geometry}

\allowdisplaybreaks

\usepackage{caption}
\usepackage{subcaption}
\usepackage{float}

\usepackage{amsfonts}
\usepackage{amssymb}
\usepackage{color}
\usepackage{multirow}
\usepackage{graphicx}
\usepackage{tcolorbox}

\newcommand{\subparagraph}{}
\usepackage[compact]{titlesec}
\titlespacing*{\section}{15pt}{1.2\baselineskip}{0.9\baselineskip}

\allowdisplaybreaks

\newcommand{\myhash}{%
	{\settoheight{\dimen0}{C}\kern-.05em\, \resizebox{!}{\dimen0}{\raisebox{\depth}{\#}}}}

\usepackage[numbers,sort&compress]{natbib}

\newcommand{\Sigmay}{{\Sigmay}_{\yv}}

\def\sigul{{\boldsymbol{\sigma}_{\text{ul}}}}
\def\sigdl{{\boldsymbol{\sigma}_{\text{dl}}}}
\def\Sigul{{\boldsymbol{\Sigma}_{\text{ul}}}}
\def\Sigdl{{\boldsymbol{\Sigma}_{\text{dl}}}}

\usepackage{multicol}

\usepackage{algorithm}
\usepackage{algpseudocode}
\usepackage{pifont}
\usepackage{varwidth}


\input{symbols.tex}

\usepackage{tikz}
\usepackage{pgfplots,tikz-3dplot}
\usetikzlibrary{intersections, pgfplots.fillbetween}
\pgfplotsset{compat=newest}
\usetikzlibrary{patterns}
\usetikzlibrary{positioning}
\usetikzlibrary{datavisualization}
\usetikzlibrary{datavisualization.formats.functions}
\usetikzlibrary{backgrounds}
\usetikzlibrary{shapes,snakes}

\usepgfplotslibrary{external} 
\usepackage{float}
\usepackage{afterpage}
\usepackage{balance}

\def\herm{{\sfH}}

\def\cg{{\clC\clN}} 
\newcommand{\normd}[1]{{\left\vert\kern-0.25ex\left\vert\kern-0.25ex\left\vert #1 
		\right\vert\kern-0.25ex\right\vert\kern-0.25ex\right\vert}}

\input{macros}

\newcommand{\Cdlk}{\Sigmam_k}
\newcommand{\Cdlcirck}{\mathring{\Sigmam}_k}

\newcommand{\ful}{f_{\rm ul}}
\newcommand{\fdl}{f_{\rm dl}}

\newcommand{\Tdl}{T_{\rm dl}}

\newcommand{\heff}{{\check{\bfh}}_{\rm eff}}
\newcommand{\heffk}{{\check{\bfh}}_{\rm eff}^{(k)}}

\def\Sigul{{\boldsymbol{\Sigma}_{\text{ul}}}}
\def\Sigdl{{\boldsymbol{\Sigma}_{\text{dl}}}}

\newcommand{\hul}{\check{\bfh}_{\rm ul}}

\newcommand{\RNum}[1]{\uppercase\expandafter{\romannumeral #1\relax}}

\def\gammam{\boldsymbol{\gamma}}

\title{Uplink-Downlink Channel Covariance Transformations and Precoding Design for FDD Massive MIMO}

\author{ Mahdi Barzegar Khalilsarai, Yi Song, Tianyu Yang, Saeid Haghighatshoar,   and Giuseppe Caire  
\thanks{The authors are with the Communications and Information Theory Group (CommIT), Technische Universit\"{a}t Berlin (\{m.barzegarkhalilsarai, yi.song, tianyu.yang, saeid.haghighatshoar,  caire\}@tu-berlin.de).}
}

\begin{document}

\maketitle

\newpage

\def\ful{f_\text{ul}}
\def\fdl{f_\text{dl}}
\def\asfc{\scrC}
\def\asful{\scrC_\text{ul}}
\def\asfdl{\scrC_\text{dl}}

%
%

\begin{abstract}
	A large majority of cellular networks deployed today make use of Frequency Division Duplexing (FDD) where, in contrast with Time Division Duplexing (TDD), the  channel reciprocity does not hold  and explicit downlink (DL) probing and uplink (UL) feedback are required in order to achieve spatial multiplexing gain. 
In order to support massive MIMO, i.e., a very large number of antennas at the base station (BS) side, the overhead incurred by conventional DL probing and UL feedback schemes scales linearly with the number of BS antennas and,  therefore, may be very large.
	In this paper, we present a new approach to achieve a very competitive tradeoff between spatial multiplexing gain and probing-feedback overhead in such systems. Our approach is based on two novel methods: (i)\,an efficient regularization technique  based on Deep Neural Networks (DNN) that learns the Angular Spread Function (ASF) of users channels and permits to estimate the DL covariance matrix  from the noisy i.i.d. channel observations obtained freely via UL pilots (UL-DL covariance transformation), (ii)\,a novel ``\textit{sparsifying precoding}'' technique that uses the estimated DL covariance matrix from  (i) and imposes a \textit{controlled sparsity} on the DL channel such that given any assigned DL pilot dimension, it is able to find an optimal sparsity level and a corresponding sparsifying precoder for which the ``\textit{effective}'' channel vectors  after sparsification can be estimated at the BS with a low mean-square error. 
	We compare our proposed DNN-based method in (i) with other methods  in the literature via numerical simulations  and show that it yields a very competitive performance.  We also compare our sparsifying precoder in (ii)  with the state-of-the-art statistical beamforming methods under the assumption that  those methods also have access to the covariance knowledge in the DL and show that our method yields higher spectral efficiency since it uses in  addition the instantaneous channel information after sparsification. 
\end{abstract}


\section{Introduction}\label{sec:intro}
Massive Multiple-Input Multiple-Output (MIMO) is a variation of conventional multi-user MIMO, where base station (BS) has a much larger number of antennas (or antenna ports) $M \gg 1$, and is considered to be a key technology for the next generation of wireless networks \cite{larsson2014massive}. Large number of antennas permits to multiplex $K \gg 1$ data streams over the spatial domain to serve $K$ users and guarantees significant advantages such as energy efficiency due to a large beamforming gain, reduced inter-cell interference, and simple user scheduling and rate adaptation due to the well-known channel hardening phenomenon \cite{larsson2014massive}.
To achieve such benefits, especially in the a downlink (DL) scenario we are interested in this paper, the BS needs to learn the channel vectors  of $K$ users to $M$ BS antennas in the DL.  With {\em Time Division Duplexing} (TDD), due to uplink-downlink (UL-DL) channel reciprocity which holds under suitable calibration \cite{marzetta2006much, Huh11}, this can be done via transmitting mutually orthogonal  pilots from the users only in the UL. Unfortunately, channel reciprocity does not hold in  {\em Frequency Division Duplexing} (FDD) since UL and DL channels lie on disjoint and far-separated frequency bands. Consequently,  the only way to learn the DL channel is to devote  a fraction of  \textit{resource elements}  (RE) \cite{lin20185g}  to estimate the DL channel of $K$ users by transmitting DL pilots and then UL feedback. Conventional DL training consists of the transmission of an $M \times \Tdl$ pilot matrix over $\Tdl$ REs, such that $\Tdl \geq M$ to  permit each user to estimate its own DL channel vector. Then, the users feed  their estimated channels back to the BS via the UL channel. Although this method works quite well for conventional MIMO systems with moderately small number of antennas $M$, it is quite inefficient in massive MIMO since $M\gg 1$ and full training of DL channel wastes at least  $\Tdl=M$ REs which may exhaust or be even larger than the whole REs available in the DL. This feedback  bottleneck makes implementing  massive MIMO in FDD quite challenging. 

\begin{figure}[t]
	\centering
	\includegraphics[scale=0.35]{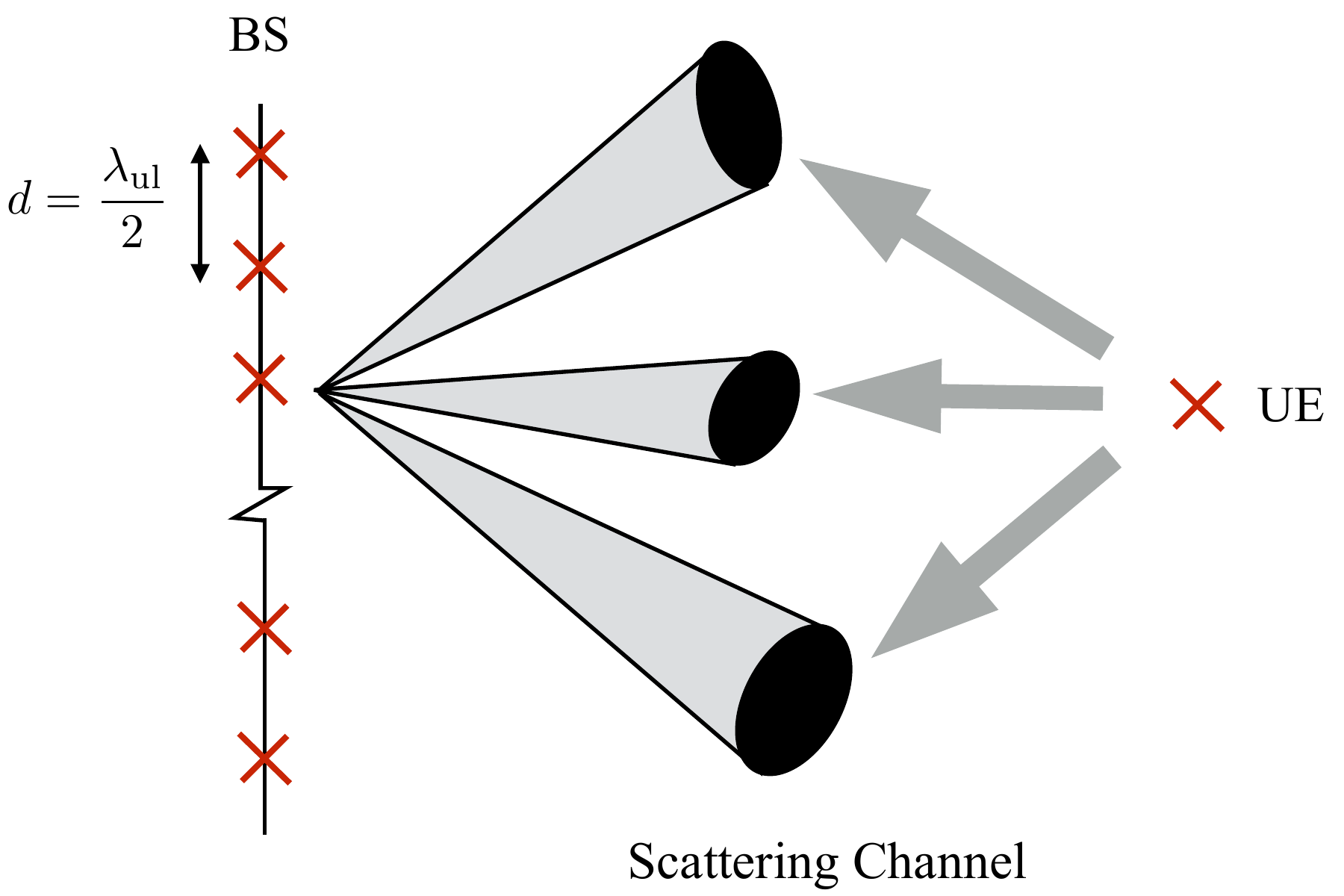}
	\caption{Sparse scattering channel between a generic user and the BS. In this example, the channel consists of 3 large scatterers reflecting the power of the user to the BS array.}
	\label{fig:scat_chan}
\end{figure}

To overcome this bottleneck, several works have been proposed to reduce DL training and feedback overhead using the sparse structure of the channel in the angle-of-arrival (AoA) domain. 
This sparsity arises due to the fact that FDD systems in 5G will be mainly used for 
large cells (while TDD for smaller denser cells) with tower-mounted base stations \cite{shepard2012argos}, where  the communication between the users and the BS occurs through a sparse cluster of scatterers with limited angular support (see, e.g.,  Fig.\,\ref{fig:scat_chan}).  
As a result, the effective dimension of the channel $s$  is much less than $M$. Building on this idea, these works proposed using compressed DL pilots and {\em Compressed Sensing} (CS)-based channel recovery techniques \cite{rao2014distributed,gao2015spatially,kuo2012compressive}. 
From standard results in CS, these method require a pilot dimension of order $\Tdl = O(s \log M)$, which may be much less than $M$ for very sparse channels $(s \ll M)$. Although the assumption $s \ll M$ may be met in very sparse Line-of-Sight (LoS) scattering scenarios, it is not fulfilled when the propagation occurs through diffuse clusters of moderately large AoA support size because in those cases the effective dimension $s$ also scales like $s=\alpha M$ where $\alpha$ is proportional to the angular width of the scatterer. In those cases, CS methods incur a pilot dimension overhead of $\Tdl = O(\alpha M \log M)$ which still grows proportionally to $M$. 
Overall, although the CS methods are able to exploit the sparsity of the channel they are still at the mercy of sparsity induced by the propagation environment since they are unable to ``shape the channel sparsity'' as desired.

\subsection{Contribution}
In this paper we propose an efficient scheme for realizing FDD massive MIMO.  The  fundamental assumption we make  to achieve this goal is that, the channel is a Gaussian process with an
\textit{Angular Spread Function} (ASF) that remains the same for UL and DL. In particular, this implies that although the UL-DL reciprocity may not hold for the instantaneous channel vectors, some sort of statistical reciprocity still holds as the ASF does not vary between  UL and  DL. We can summarize our contributions in this paper as follows.

\noindent {\bf (i) A Novel UL-DL Covariance Transformation Using Deep Neural Networks. }
We use the reciprocity of the ASF in UL and DL to estimate the DL covariance matrix from the observation of noisy channel vectors freely available through the UL pilots. The ASF reciprocity  was also assumed implicitly  in several past works \cite{hochwald2001adapting, han2010potential, decurninge2015channel} by considering only ASFs with spike discrete components. 
Recent works \cite{haghighatshoar2018multi, miretti2018fdd, cavalcante2018error} generalize these results to arbitrary ASFs consisting of both discrete spike and also continuous components, and show that such Uplink-Downlink covariance transformation (UDCT) is still feasible for sufficiently large number of BS antennas $M$ although establishing the stability of UDCT is quite challenging when only a noisy estimate of the UL covariance  matrix is available. 

One of the aims of this paper is to improve the performance of  UDCT by exploiting the natural group sparsity structure of the ASF. Unfortunately, designing good regularization methods that promote the group-sparsity of the ASF (see, e.g., Fig.\,\ref{fig:scat_chan}) is not trivial. In particular, none of the structured group-sparse estimators widely adopted in sparse signal recovery literature are applicable here \cite{baraniuk2010model, eldar2009block}.  In this paper, we bridge this gap by using Deep Neural Networks (DNNs). We train DNNs with suitable training data corresponding to group-sparse ASFs and let them learn/capture the notion of group-sparsity. Once the DNN is suitably trained, we use it as a black-box algorithm to estimate the ASF, thus, the DL covariance matrix, from the noisy channel vectors gathered in the UL. We illustrate via numerical simulations that our proposed DNN-based method is quite strong and recovers the DL covariance matrix much better than other methods.
\vspace{1mm}

\noindent {\bf (ii)\,A Novel Sparsifying Precoder.} We use the  estimated DL covariance matrix from (i) to design a {\em sparsifying precoder}, such that the following two criteria are met: 
(a)\,the reduced-dim channel vectors of all the users after sparsification are sparse enough such that they  are stably estimated with a very low error and (b)\,the dimension of the channel after sparsification is kept as large as possible in order to increase the rank of the effective channel matrix of the users after sparsification such  that large number of users $K$ can be served.
We pose this  as a Mixed Integer Linear Program (MILP) and solve it via off-the-shelf MILP solvers.

It is also worthwhile to mention that compared with  CS methods, which are at the mercy of the sparsity of the channel due to the propagation,  our proposed method is able to shape the sparsity of the channel completely  flexibly depending on number of REs $\Tdl$ available for channel estimation.
We illustrate via numerical simulations that the proposed method has an excellent performance much superior to other methods such as statistical beamforming, which also use the knowledge of the DL covariance matrix.

\subsection{Notation}
We denote vectors/matrices with small/large boldface letters (e.g., $\bfx$/$\bfX$), and sets with calligraphic letters (e.g., $\clX$).
We use the $i$-th element of a vector $\bfx$ with $[\bfx]_i$ and the $(i,j)$-th element of a matrix $\bfX$ with $[\bfX]_{i,j}$. For an integer $k$, we use the short-hand notation $[k]$ for $\{1,\dots, k\}$.

\section{Proposed ASF Estimation and Uplink-Downlink Covariance Transformation}\label{sec:dnn}
In this section, we explain our proposed method for UDCT using DNNs. For the sake of completeness, we first provide a summary of UDCT problem for the Uniform Linear Array (ULA) we consider in this paper (see, Fig.\,\ref{fig:scat_chan}).
%
Let us consider a generic use and let us denote the ASF of this user by $\gamma(\xi)$ where $\xi=\sin(\theta) \in [-1,1]$ denotes a parametrization of the AoA $\theta\in[-\frac{\pi}{2}, \frac{\pi}{2}]$ and where $\gamma(\xi)$ is  density of the received signal power at the AoA $\xi$. The covariance matrix of the channel vector of this user at UL/DL carrier frequency $f \in \{\ful, \fdl\}$ is given by \cite{haghighatshoar2018multi}
\begin{align}\label{cov_cont}
\Sigmam(f)=\int_{-1}^1 \gamma(\xi) \bfa(\xi, f) \bfa(\xi,f)^\herm   d\xi,
\end{align}
where $\bfa(\xi,f)$ denotes the array response vector at AoA $\xi$ at frequency $f$ given by
\begin{align}
[\bfa(\xi,f)]_k=e^{j (k-1) \pi d \frac{f}{c_0} }, k \in [M],
\end{align}
where $d$ denotes the antenna spacing and where $c_0$ is the speed of light. 
Note that in \eqref{cov_cont} we assumed implicitly that the ASF $\gamma(\xi)$ is the same in both UL and DL frequency range. As explained before, this provides some sort of statistical UL-DL channel reciprocity for FDD (in contrast with the instantaneous channel reciprocity which may not hold). 
With this notation, can pose the UDCT problem as follows.

\noindent {\bf UDCT Problem:} 
{\em Given the UL covariance matrix or an estimate thereof $\Sigul:=\Sigmam(f_\text{ul})$, find the DL covariance matrix $\Sigdl:=\Sigmam(f_\text{dl})$.}  \hfill $\lozenge$

For the ULA, we can gain  a better understanding of UDCT by looking at the Fourier coefficients of the ASF $\gamma(\xi)$. We first assume that the array has the standard half wavelength  spacing in the UL, namely, $d=\frac{\lambda_\text{ul}}{2}$ where $\lambda_\text{ul}=\frac{c_0}{\ful}$ denotes the wavelength at UL carrier frequency. Then, it is not difficult to show that $\Sigul$ is a Toeplitz matrix whose first column is given by $\sigul(\gamma) \in \bC^M$ where
\begin{align}\label{UL_proj}
[\sigul(\gamma)]_k=\int_{-1}^1 \gamma(\xi) e^{j (k-1) \pi \xi} d\xi,\ \  k\in [M],
\end{align}
denotes the $(k-1)$-th Fourier coefficient of $\gamma(\xi)$. Similarly,  $\Sigdl$ is also a Toeplitz matrix with first column $\sigdl(\gamma)\in \bC^M$
\begin{align}\label{DL_proj}
[\sigdl(\gamma)]_k=\int_{-1}^1 \gamma(\xi) e^{j (k-1) \pi \beta \xi} d\xi,\ \  k\in [M],
\end{align}
where $\beta=\frac{\fdl}{\ful}$ denotes the ratio between the DL and UL carrier frequencies. In current deployments of FDD systems typically $\ful>\fdl$, thus, $\beta>1$.  Therefore, in the ideal case where $\Sigul$ is known exactly, one can pose UDCT as recovering the projections of the positive functions $\gamma(\xi)$ on the DL set of harmonic functions $\clH_\text{dl}=\{e^{j (k-1) \pi \beta \xi}: k\in [M]\}$ from the knowledge of its projections on the UL harmonic functions $\clH_\text{ul}=\{e^{j (k-1) \pi  \xi}: k\in [M]\}$. 

All the UDCT algorithms in the literature implicitly or explicitly do the following: (a)\,estimate a positive function $\widehat{\gamma}(\xi)$ that has exactly (in the noiseless case) or approximately (in the noisy case) the same projections on the UL harmonic functions $\clH_\text{dl}$ as the original ASF $\gamma(\xi)$, (b)\,use the resulting estimate $\widehat{\gamma}(\xi)$ to compute the projections onto the DL set $\clH_\text{dl}$ to  recover an estimate of the DL covariance matrix. Note that even in the ideal noiseless case, the mapping from the ASF to the DL projections is a linear map from the infinite-dim space of positive functions to the finite-dim space of $M$ projections. Therefore, there are generally a large set of ASFs corresponding to a given UL projection $\sigul(\gamma)$ produced by a generic ASF $\gamma(\xi)$:
\begin{align}
\clA(\gamma):=\big \{\mu(\xi): \int_{-1}^1 \mu(\xi) e^{j (k-1)\pi \xi} d \xi=[\sigul(\gamma)]_k\big \}.
\end{align}
\def\alg{\texttt{ALG}}
All the UDCT algorithms can be seen in one way or other as simply different strategies for selecting a specific candidate $\alg(\sigul(\gamma)) \in \clA(\gamma)$ according to specific criteria. 
One such criterion is to assume that the original  ASF $\gamma$ belongs to a specific subset $\clA_0$ of structured ASFs. Also, the UDCT algorithm can written more generally as a method that produces $\alg(\sigul(\gamma)) \in \clA(\gamma) \cap \clA_0$ when fed with the UL projections $\sigul(\gamma)$. 

\def\hul{\bfh_\text{ul}}
A similar argument applies to the more general case where instead of $\Sigul$ one has access to a collection of $N$ i.i.d. noisy UL channel vectors $\bfy(s)=\hul(s)+\bfz(s)$, $s\in [N]$, gathered via UL pilot transmission where $\hul(s)$ and $\bfz(s)$ denote the UL channel vector and additive measurement noise over resource block $s \in [N]$. Then, one can compute the UL sample covariance matrix as
\begin{align}\label{ul_samp_cov}
\widehat{\Sigul} =\frac{1}{N} \sum_{s\in [N]} \bfy(s) \bfy(s)^\herm,
\end{align}
and design an algorithm $\alg$ that produces a structured ASF in $\clA_0$ whose UL covariance matrix matches $\widehat{\Sigul}$ under suitable metric. 
For example, \cite{haghighatshoar2018multi, miretti2018fdd} propose such algorithms using Non-Negative Least Squares (NNLS) and $\ell_2$-norm projections. However, none of these methods are able to capture the group-sparsity of the ASF in the AoA domain as illustrated in Fig.\,\ref{fig:scat_chan}. For example, as we will illustrate via numerical simulations, the NNLS proposed in \cite{haghighatshoar2018multi} is able to promote sparsity of ASF in the AoA domain but it does not yield necessarily group-sparse ASF. The $\ell_2$ projection method \cite{miretti2018fdd}, in contrast, is able to produce smooth ASFs but creates out-of-support components in the estimated ASF since $\ell_2$ norm is inherently unable to promote sparsity.

Overall it is generally  difficult to design suitable regularization methods that promote notion of group-sparsity of ASF we address here. In particular, none of the structured group-sparse estimators widely adopted in sparse signal recovery literature are applicable here \cite{baraniuk2010model, eldar2009block}. 
In this paper, we develop such a group-sparsity promoting  regularization  using DNNs as follows. 

\newcommand{\ch}[1]{\check{#1}}
\newcommand{\wch}[1]{\widetilde{#1}}

\noindent{\bf (a) Training Data.} We first consider a class of group-sparse ASFs $\clA_0$ that may potentially arise in practical propagation scenarios. To capture the notion of group sparsity, we assume that each  ASF in $\clA_0$ can be written as $\gamma(\xi)=\sum_{i=1}^g \kappa_i p_i(\xi)$ where $g$ denotes the number of  groups, where $p_i(\xi)$ is a normalized, i.e., $\int_{-1}^1 p_i(\xi) d\xi=1$, positive function with connected support in $[\xi_i, \xi_i+w_i]$ with $w_i$ denoting the angular width of the ASF of the $i$-th group $p_i(\xi)$, and where $\kappa_i \in [0,1]$ with $\sum_{i=1}^g \kappa_i=1$ are the normalized weights corresponding to ASFs in $g$ groups. For example, Fig.\,\ref{fig:scat_chan} corresponds to an ASF with $g=3$ groups. 

\begin{figure}
	\centering
	\includegraphics[scale=0.5]{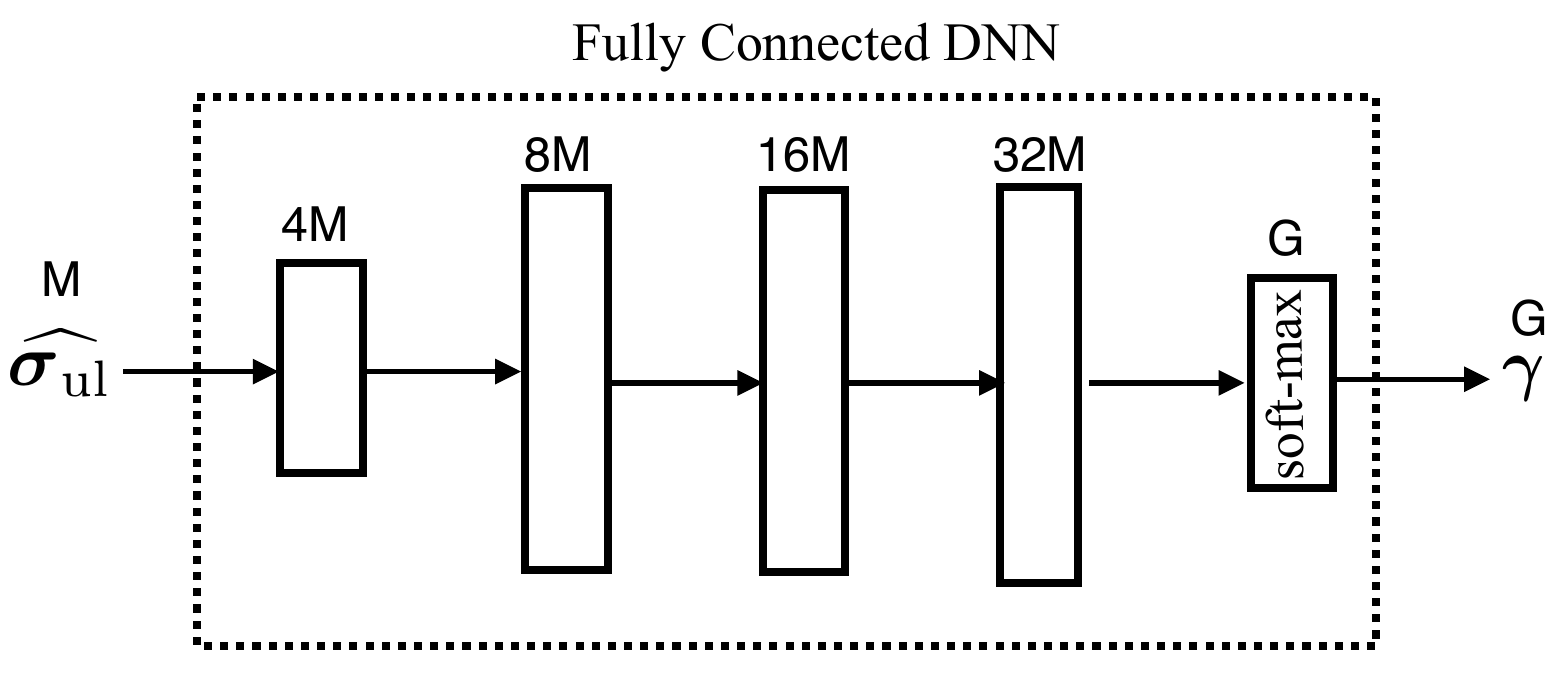}
	\caption{The structure of DNN adopted for ASF estimation. DNN consists of $5$ layers with $2M$, $4M$, $8M$,  $16M$, and $G$ neurons, where $M$ is the number of antennas and where $G$ is the ASF quantization grid size. The last layer has a  \textit{soft-max} activation function and produces positive values for $\gammam\in \bR_+^{G}$. }
	\label{DNN}
\end{figure}

 We use the ASFs in $\clA_0$ to train a DNN as follows. Given a training sample size $S$, we select $S$ ASFs from $\clA_0$ completely randomly. For each specific ASF $\gamma$ inside this training set, we compute the first row of the UL covariance matrix as \eqref{UL_proj} and produce $N$ i.i.d. noisy UL channel vectors and their corresponding sample covariance $\widehat{\Sigul}$ as in \eqref{ul_samp_cov}. Since the covariance matrices for ULA are Toeplitz, we toeplitizify $\widehat{\Sigul}$ and define the first column of the resulting Toeplitz matrix as $\widehat{\sigul}$ where
 \begin{align}\label{toeplitzif}
 [\widehat{\sigul}]_k=\frac{\sum_{i=1}^{M-k+1} [\widehat{\Sigul}]_{i,i+k}}{M-k+1}. 
 \end{align}
We  define a uniform quantization grid $\clG:=\{\xi_i: i \in [G]\}$ over the set of AoAs $[-1,1]$ of size $G \gg M$ where $\xi_i=-1+\frac{2(i-1)}{G}$ denotes the $i$-th quantization point. We also define the discrete quantization of the ASF $\gamma(\xi)$ over the grid $\clG$ as $\gammam=(\gamma(\xi_1), \dots, \gamma(\xi_G))^\transp\in \bR_+^G$. For simplicity, we always  normalize $\gammam$ to make sure that $\sum_{i=1}^G [\gammam]_i=1$. Finally, we use $(\widehat{\sigul}, \gammam)$ as input-output labeled pair for training the DNN illustrated in Fig.\,\ref{DNN}. By repeating this for all the $S$ ASFs selected for training, we obtain a collection of $S$ training samples for DNN.

\noindent{\bf (b) Supervised Learning.}  We use these $S$ training samples to train a \textit{Deep Neural Networks} (DNN). For training, we use  the widely-adopted \textit{Stochastic Gradient Descend} (SGD) with $\ell(\gammam, \widehat{\gammam})=\|\gammam - \widehat{\gammam}\|_1$ as the loss function between the true $\gammam$ and the estimate $\widehat{\gammam}$ generated by the network. DNNs have recently been of tremendous importance in Machine Learning and also in many applications in wireless communications, such as signal detection, channel encoding \cite{farsad2018deep}, and decoding \cite{nachmani2018deep}.


\noindent{\bf (c) Structure of DNN.}
One of the important factors affecting the performance of the ASF estimation using DNNs is the structure of the DNN consisting of the number of layers, the number of neurons in each layer, and  the activation function of each layer. In this paper, we use a fully-connected network illustrated in Fig.\,\ref{DNN}, with $5$ layers consisting of $2M$, $4M$, $8M$, $16M$, and $G$ neurons, respectively, where $M$ denotes the number of antennas, where the number of neurons $G$ in the last layer corresponds to the grid size we are adopting for ASF quantization. The activation function of the $4$ initial layers is the RelU function $x \mapsto \max\{x,0\}$. For the last layer we use the \textit{soft-max} activation function, which for an input vectors $(x_1, \dots, x_G) ^\transp$ in the input produces the output as $(x_1, \dots, x_{G})^\transp \mapsto \frac{(e^{x_1}, \dots, e^{x_{G}})^\transp}{\sum_{j=1}^{G} e^{x_j}}$. Note that the summation of the elements produced by soft-max layer is always $1$, which produces a normalized $\gammam$, i.e., $\sum_{i=1}^G\gamma_i=1$, as desired.

\noindent{\bf (d) UDCT using DNN.} Once the DNN was suitably trained we use it as a black-box algorithm for ASF estimation and UDCT. More specifically, given $N$ i.i.d. noisy channel vectors received in UL, we compute the UL sample covariance matrix as in \eqref{ul_samp_cov} and $\widehat{\sigul}$ as in \eqref{ul_samp_cov}. Then, we feed $\widehat{\sigul}$ to the trained DNN and obtain an estimate of the quantized ASF $\widehat{\gammam}\in \bR_+^G$ where $G$ is the grid size. Then, we build an estimate of the original continuous ASF $\widehat{\gamma}(\xi)$ by triangular interpolation. Finally, we use $\widehat{\gamma}(\xi)$ to compute the first column of the  DL Toeplitz covariance matrix as in \eqref{DL_proj}. 
Note that DNNs have the fundamental advantage that one does not need to run any iterative algorithm as in \cite{haghighatshoar2018multi, miretti2018fdd}, which may require many iteration to converge; instead one immediately computes the estimate by straightforward calculations through the network, which can be done even in parallel to obtain a tremendous speed-up.

\section{Sparsification Precoding}\label{sec:ch_sparsification}
In this section, we provide an step-by-step method to design the sparsifying precoder for a collection of $K$ users based on their estimated DL covariance matrix denoted by $\Cdlk$, $k \in [M]$, where for simplicity we drop the label `dl'. 
An essential  requirement for the our method is  the existence of the \textit{common eigen-vector property} for the array and in particular for the set of $K$ DL covariance matrices $\Cdlk$, $k \in [K]$. In brief, this property is satisfied when the set of all the DL covariance matrices produced by the array is, at least asymptotically for large $M$, diagonalizable in the same basis. Fortunately, this assumption is met for Uniform Linear/Planar Array (ULA/UPA) widely adopted in wireless applications since the resulting covariance matrices have Toeplitz/Block-Toeplitz structure. And, it is well-known that, asymptotically for large $M$, all such matrices are diagonalizable with appropriate DFT (Discrete Fourier Transform) matrices, thus, common eigen-vector property. 

In the following, we provide a detailed derivation of the this property for the ULA.
We first calculate the circulant approximation of the estimated DL covariance matrices of all users, so that they all share the same set of eigen-vectors, or equivalently virtual beam-space representation.\footnote{Recall that circulant matrices are diagonalizable with the orthogonal DFT matrix.} The circulant approximation of large Toeplitz matrices imposes an small error as a result of the application of Szeg\"o
Theorem \cite{adhikary2013joint}. 
Let $\Cdlk$, $k\in[K]$, be as before and define the diagonal matrices $\mathring{\Lambdam}_k = \text{diag}(\Fm^\herm \Cdlk \Fm)$ for $k \in [K]$. 
There are several ways to define a circulant approximation \cite{zhu2017asymptotic}, among which we choose  
$\Cdlcirck= \Fm \mathring{\Lambdam}_k\Fm^\herm$.
According to Szeg\"o's theorem, for large $M$, $\mathring{\Lambdam}_k$ converges to the diagonal eigenvalue matrix $\Lambdam_k$ of $\Cdlk$, i.e. $\mathring{\Lambdam}_k \rightarrow \Lambdam_k$ as $M\rightarrow \infty$. 
This shows that, with a small error, we can find a set of common eigenvectors for all the 
estimated DL covariance matrices $\Cdlk$, $k \in [K]$. As a consequence, the DL channel covariance of each user $k\in [K]$ is characterized simply 
by a vector of eigenvalues $\lambdav_{k} \in \bR^M$, with $m$-th element 
$\lambda^{(k)}_m = [\mathring{\Lambdam}^{(k)}]_{m,m}$. 
In addition, the DFT matrix whose $(m,n)$-th entry is given by $[\Fm]_{m,n}=\frac{1}{\sqrt{M}} e^{-j2\pi \frac{mn}{M}},~m,n\in [M]$, forms a unitary basis 
for (approximately) expressing any user channel vector via an (approximated) Karhunen-Loeve expansion. In particular, 
letting $\fv_n:=[\Fm]_{\cdot,n}$ denote the $n$\textsuperscript{th} column of $\Fm$, we can express the DL channel vector of user $k$ 
as 
\begin{equation} \label{eq:approximate-KL}
\hv^{(k)} \approx  \sum_{m=0}^{M-1} g_{m}^{(k)} \sqrt{\lambda_{m}^{(k)}} \fv_n 
\end{equation}
where $g_{m}^{(k)} \sim \cg (0,1) $. Here, we wish to design the precoder $\Bm$ such that the support of the effective channels $\heffk = \Bm \hv^{(k)}$ is not 
larger than $\Tdl$, such that all users have a chance of being served. 
Let $\Hm = \Lm \odot \bG\in \bC^{M\times K}$ denote the matrix of DL channel coefficients expressed in the DFT basis (\ref{eq:approximate-KL}),  
in which each column of $\Hm$ represents the coefficients vector of a user,  where $\Lm$ is a $M \times K$ matrix with elements 
$[\Lm]_{m,k} = \sqrt{\lambda_m^{(k)}}$,  where $\bG \in \bC^{M\times K}$ has i.i.d. elements  $[\bG]_{m,k} = g_{m}^{(k)}$, 
and where $\odot$ denotes the Hadamard (elementwise) product. 

\begin{figure}[t]
	\centering
	\begin{subfigure}[b]{.24\textwidth}
		\centering
		\includegraphics[width=1\linewidth]{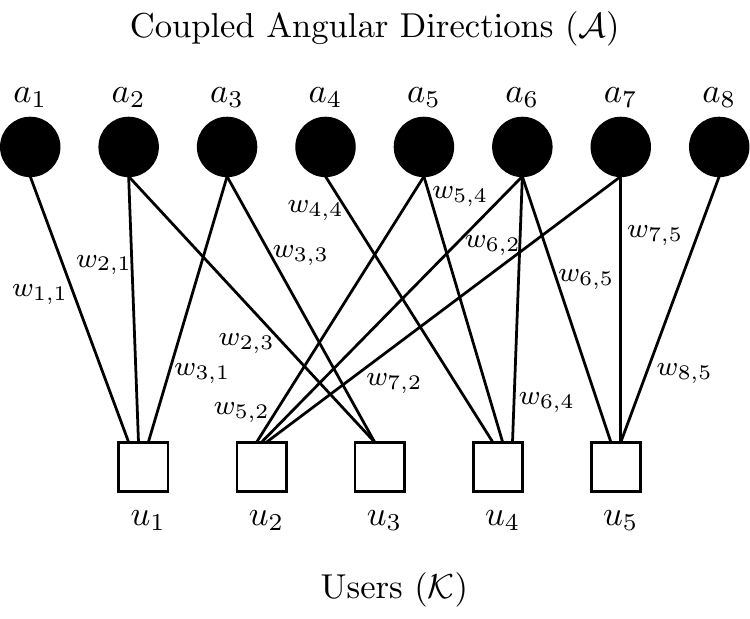}
		\caption{}
		\label{fig:bi_graph}
	\end{subfigure}
	\begin{subfigure}[b]{.24\textwidth}
		\centering
		\includegraphics[width=0.6\linewidth]{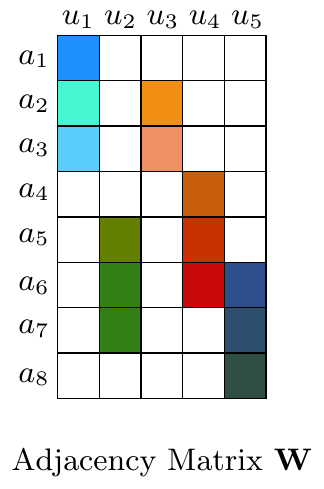}
		\caption{}
		\label{fig:adj_mat}
	\end{subfigure}
	\caption{\small (a) An example of a bipartite graph $\Lc$. (b) The corresponding weighted adjacency matrix $\Wm$.}
	\label{fig:test}
\end{figure}

To design the sparsifying precoder, we first illustrate the joint user-beam association of all $K$ users by a graphical model. Let $\Am = [\Lm]$ denote a one-bit thresholded version of $\Lm$, such that $[\Am]_{m,k}=1$ if $\lambda_m^{(k)} > \epsilon$, where $\epsilon > 0$ is a suitable
small threshold, used to identify the components that are significantly larger than 0 from the ``almost zero'' ones,   and consider the $M \times K$ bipartite 
graph $\Lc = \left(\Ac,\Kc,\Ec\right)$ with adjacency matrix $\Am$ and weights $w_{m,k} = \lambda_m^{(k)}$  on the edges $(m,k) \in \Ec$. An example of the bipartite graph $\Lc$ and its corresponding weighted adjacency matrix $\Wm$ is illustrated in Figs.\,\ref{fig:bi_graph} and \ref{fig:adj_mat}.
Given a pilot dimension $\Tdl$, our goal consists in selecting a subgraph $\Lc'=\left(\Ac',\Kc',\Ec'\right)$ of $\Lc$ in which each node on either side of the graph has a degree at least 1 and such that 
\begin{enumerate}
	\item For all $k \in \Kc'$ we have $\text{deg}_{\Lc'}(k) \le \Tdl$, where $\text{deg}_{\Lc'}$ denotes the degree of a node in the selected subgraph.
	\item The sum of weights of the edges adjacent to any node $k \in \Kc'$ in the subgraph $\Lc'$ is greater than a threshold, i.e. $\sum_{m\in \Nc_{\Lc'} (k)} w_{m,k} \ge P_0, ~\forall k\in \Kc'$, where $\Nc_{\Lc'}(k)$ denotes the set of neighbors in  $\Lc'$ of node $k$. 
	\item The channel matrix $\Hm_{\Ac',\Kc'}$ obtained from $\Hm$ by selecting $a \in \Ac'$ (referred to as ``selected beam 
	directions") and $k\in \Kc'$ (referred to as ``selected users") has large rank.
\end{enumerate}    
The first criterion enables the stable estimation of the effective channel of any selected user with only $\Tdl$ common pilot dimensions 
and $\Tdl$ complex symbols of feedback per selected user. 
The second criterion makes sure that the effective channel strength of any selected user is greater than a certain desired threshold. 
The third criterion is motivated by the fact that the DL multiplexing gain is given by ${\rm rank} ( \Hm_{\Ac',\Kc'} ) \times \max \{ 0 , 1 - \Tdl/T\}$, where $T$ denotes the whole number of REs, 
and it is obtained by serving a number of users equal to  the rank of the effective channel matrix. 

In \cite{khalilsarai2018fdd} we show that the rank of $\Hm$ is given, with probability 1, by the size of the largest intersection submatrix whose associated bipartite graph contains a perfect matching.\footnote{A matching is a set of edges of a graph without common vertices.} With this observation, we can formulate the problem mentioned above as follows. Let $\Mc (\Ac',\Kc')$ denote a matching of the subgraph $\Lc' (\Ac',\Kc',\Ec')$ 
of the bipartite graph $\Lc (\Ac,\Kc,\Ec)$. Find the solution of the following optimization problem:
\begin{subequations}\label{eq:my_opt_1}
	\begin{align}
	& \underset{\Ac'\subseteq \Ac, \Kc' \subseteq \Kc}{\text{maximize}} && \left\vert \Mc\left(\Ac',\Kc'\right)\right\vert    \label{eq:my_opt_1-one}  \\
	& \text{subject to} &&  \text{deg}_{\Lc'} (k) \le \Tdl~ \forall k\in \Kc',  \label{eq:my_opt_1-two} \\
	& ~ &&  \hspace{-4mm} \sum_{a\in \Nc_{\Lc'} (k) } w_{a,k}  \ge P_0, ~ \forall k\in \Kc'. \label{eq:my_opt_1-three}
	\end{align}
\end{subequations} 
It turns out that, this problem can be transformed into the following equivalent mixed integer linear program (MILP), denoted by $\Pc_{\text{MILP}}$ \cite{khalilsarai2018fdd}:
\begin{subequations}\label{opt:P_MILP}
	{	\small
		\begin{align}
		\underset{x_m,y_k,z_{m,k} }{\text{maximize}}  & ~~\sum_{m\in \Ac} \sum_{k \in \Kc} z_{m,k}   \label{eq:obj_1} + \epsilon \sum_{m \in \Ac} x_m \\
		\text{subject to}  &  ~~~~z_{m,k} \le [\Am]_{m,k} ~~\forall m\in\Ac,k\in \Kc, \label{eq:one}\\
		~~ & ~~~~ \sum_{k\in \Kc}z_{m,k} \le x_m ~~ \forall m\in \Ac, \label{eq:two}\\ 
		~~ & ~~~~ \sum_{m\in \Ac}z_{m,k} \le y_k ~~ \forall k\in \Kc, \label{eq:three} \\
		~~ & ~~~~ \sum_{m\in \Ac} [\Am]_{m,k} x_m \le \Tdl y_k + M (1-y_k) ~~\forall k\in \Kc, \label{eq:four}\\
		~~ & ~~~~  P_0 \, y_k \le \sum_{m \in \Ac} [\Wm]_{m,k} x_m ~~\forall k \in \Kc, \label{eq:five} \\
		~~ & ~~~~  x_m \le \sum_{k\in \Kc} [\Am]_{m,k} y_k ~~ \forall m\in \Ac, \label{eq:six}  \\
		~~ & ~~~~  x_m, y_k \in \{0,1\} ~~\forall a\in \Ac,k\in \Kc,\\
		~~ & ~~~~  z_{m,k} \in [0,1] ~~\forall m\in \Ac,k\in \Kc,
		\end{align}
	}
\end{subequations}
The binary variable $x_m$ in \eqref{opt:P_MILP} determines whether beam $m\in \Ac$ is selected or not, i.e., $x_m=1$ if and only if the $m$-th beam is selected. Similarly, $y_k$ controls the selection of user $k$ for channel estimation (and eventually serving). For a given set of user DL covariance matrices, we denote by $\Bc=\{m:x_m^\ast=1\} = \{m_1,m_2,\ldots, m_{M'}\}$ the set of selected beams directions of cardinality $|\Bc| = M'$ and 
by $\Kc = \{k: y^\ast_k = 1\}$ the set of selected users of cardinality $|\Kc| = K'$, where $\left\{x_m^\ast \right\}_{m=1}^M$ and $\left\{y_k^\ast \right\}_{k=1}^K$ are solutions to $\Pc_{\text{MILP}}$. 

The desired sparsifying precoding matrix $\Bm$  is finally  obtained 
as $\Bm = \Fm_{\Bc}^\herm$, where $\Fm_{\Bc} = [\fv_{m_1}, \ldots, \fv_{m_{M'}}]$ 
and $\fv_{m}$ denotes the $m$-th column of the $M \times M$ unitary DFT matrix $\Fm$.  Using (\ref{eq:approximate-KL}), the effective DL channel vectors take on the form
\begin{equation} \label{effch}
\heff^{(k)}  = \Bm \sum_{m \in \Sc_k} g_m^{(k)} \sqrt{\lambda_m^{(k)}} \fv_m =  \sum_{i : m_i \in \Bc \cap \Sc_k}  \sqrt{\lambda_{m_i}^{(k)}} g_{m_i}^{(k)} \uv_i,
\end{equation}
where $\Sc_k$ is the the support of $\hv^{(k)}$ in the DFT domain and where $\uv_i$ denotes a $M' \times 1$ vector with all zero components but a single ``1'' in the $i$-th position. With this construction, the number of non-identically zero coefficients for each user $k$ are $|\Bc \cap \Sc_k| \leq \Tdl$ and their 
positions (encoded in the vectors $\uv_i$ in (\ref{effch}))  are known to the BS. 
Hence, the effective channel vectors can be estimated from the $\Tdl$-dimensional DL pilot observation with an estimation MSE
that vanishes as $1/\SNR$ \cite{khalilsarai2018fdd} by increasing the Signal-to-Noise Ration (SNR). With the above precoding, we have $\Bm \Bm^\herm = \Id_{M'}$. Furthermore, 
we can choose the DL pilot matrix $\Psim$ to be proportional to a random unitary matrix of dimension $\Tdl \times M'$, such that 
$\Psim \Psim^\herm = P_{\rm dl}  \Id_{\Tdl}$. After DL pilot transmission,  all users $k\in [K]$ send back their noisy observations $\yv^{(k)}$ to the BS via analog unquantized feedback. The BS, having an estimate of the DL channel covariance, performs linear MMSE estimation to obtain an estimate of the DL channel.

We concclude by describing the channel precoding step. Let $\widehat{\bfH}_{{\text{eff}}}= [ \widehat{{\bfh}}_{{\text{eff}}}^{(1)}, \ldots, \widehat{{\bfh}}_{{\text{eff}}}^{(K')}]$ be the matrix 
of the estimated effective DL channels for the selected users. We consider the ZF beamforming matrix $\Vm$ given by the column-normalized 
version of the Moore-Penrose pseudoinverse of the estimated channel matrix, i.e., $\Vm = 
\left(\widehat{\bfH}_{{\text{eff}}} \right)^\dagger \Gm^{1/2}$, where $\left(\widehat{\bfH}_{{\text{eff}}} \right)^\dagger = 
\widehat{\bfH}_{{\text{eff}}}  \left ( \widehat{\bfH}_{{\text{eff}}}^\herm \widehat{\bfH}_{{\text{eff}}} \right )^{-1}$ and $\Gm$ is a diagonal matrix that makes the columns of 
$\Vm$ to have unit norm. A channel use of the DL precoded data transmission phase at the $k$-th user receiver takes on the form
$y^{(k)} = \left ( \hv^{(k)} \right )^\herm \Bm^\herm \Vm \Pm^{1/2} \bfd + n^{(k)}$, 
where $\bfd \in \bC^{K' \times 1}$ is a vector of unit-energy user data symbols and $\Pm$ is a diagonal matrix defining the 
power allocation to the DL data streams. The transmit power constraint is given by $\trace( \Bm^\herm \Vm \Pm \Vm^\herm \Bm ) = \trace ( \Vm^\herm \Vm \Pm ) = \trace (\Pm) = P_{\rm dl}$,
where we used $\Bm \Bm^\herm = \Id_{M'}$ and the fact that $\Vm^\herm \Vm$ has unit diagonal elements by construction. 
For simulation results,  we use the simple uniform power allocation $P_k = P_{\rm dl}/K'$ to each $k$-th user data stream. The received symbol at user $k$ receiver is given by  
\begin{align}
y^{(k)} = b_{k,k} d_k + \sum_{k' \neq k} b_{k,k'} d_{k'}   +  n^{(k)},
\end{align} 
where the coefficients $(b_{k,1}, \ldots, b_{k,K'})$ are given by the elements of the $1 \times K'$ row vector
$\left ( \hv^{(k)} \right )^\herm \Bm^\herm \Vm \Pm^{1/2}$. Of course, in the presence of an accurate channel 
estimation we expect that $b_{k,k} \approx \sqrt{G_k P_k}$ and $b_{k,k'} \approx 0$ for $k' \neq k$. To calculate achievable sum-rate, we assume that all coefficients 
$(b_{k,1}, \ldots, b_{k,K'})$ are known to the corresponding receiver $k$. Including the DL training overhead, this yields the rate expression
{\small
	\begin{equation}\label{eq:rate_ub}
	R_{\rm sum}  = \left (1 - \frac{\Tdl}{T} \right ) \sum_{k \in \Kc} \bE \left [ \log \left ( 1 + \frac{\left|b_{k,k}\right|^2}{1 + \sum_{k' \neq k} \left|b_{k,k'} \right|^2} \right ) \right ].
	\end{equation}}

\section{Simulation Results}
In this section, we perform numerical simulations to assess the performance of our proposed method. 
For simulations, we consider IMT-FDD band as in LTE standard \cite{LTEbands} with a UL band $[1920,1980]$\,MHz and a DL band $[2110,2170]$\,MHz. Thus, we use $f_\text{ul}=1950$ and $f_\text{dl}=2140$, and set the UDCT parameter $\beta$ in \eqref{DL_proj} to $\beta=\frac{\fdl}{\ful}\approx 1.1>1$. 
We also set the number of BS antennas to $M=256$. 

\subsection{ASF Estimation and UL-DL Covariance Transformation}
In this section, we perform numerical simulations to compare  a DNN trained with group-sparse ASFs with NNLS method in \cite{haghighatshoar2018multi} and $\ell_2$ projection method in \cite{ miretti2018fdd}.

\noindent{\bf ASF Estimation.} We assume that each ASF in the set of feasible group-sparse ASFs consists of two groups $g=2$ where  the ASF corresponding to each group is a uniform distribution with a center randomly selected in the range of AoAs $[-1,1]$ and  with a random width of size at most  $w_{\text{max}}=0.4$. We consider a SNR of 20\,dB for noisy UL channel vectors used for training and set the sampling ratio to $\frac{N}{M}=2$ (see, e.g., \eqref{ul_samp_cov}).
We refer to Section \ref{sec:dnn} for a more detailed discussion of the data set used for training the DNN.  We build a training data set of size $S=10000$ where we use 80\% of this data set for training and the remaining 20\% for validation.

\begin{figure}[t]
	\centering
	\includegraphics[width=0.475\textwidth]{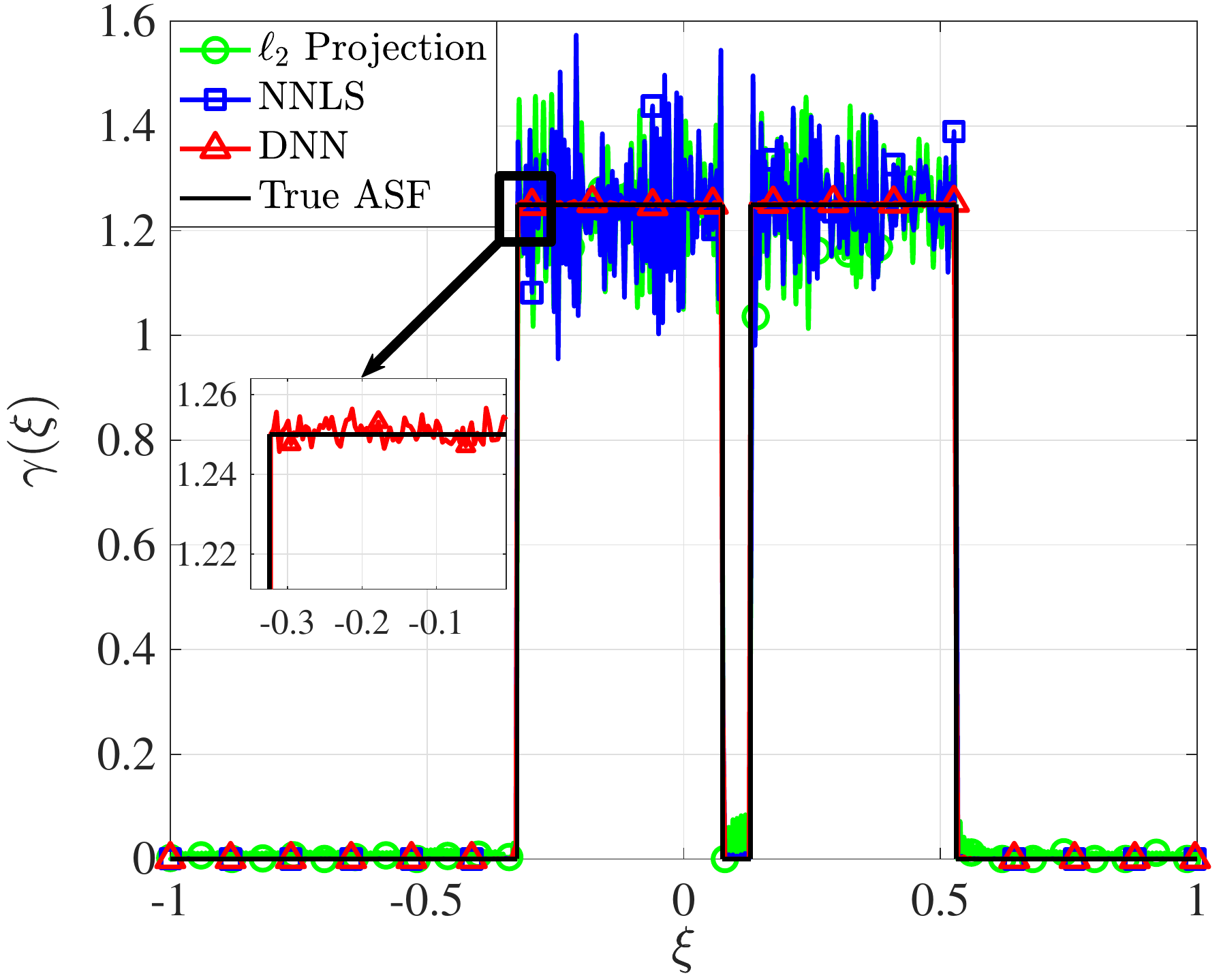}
	\caption{Comparison of performance of ASF estimation for DNN, NNLS, and $\ell_2$ Projection method.}
	\label{fig:ASF}
\end{figure}

Fig.\,\ref{fig:ASF} illustrates the simulation results for ASF estimation. It is seen that DNN performs much better than NNLS and $\ell_2$ projection. NNLS estimates the support  very well but is quite spiky  over the support. In contrast, $\ell_2$ projection method is quite smooth over the support but produces considerable off-support elements.

\noindent{\bf UL-DL Covariance Transformation.}
In this part, we compare the UDCT performance of our proposed DNN with that of  NNLS and $\ell_2$ projection method  for different sampling ratios $\frac{N}{M}\in \{1, \dots, 8\}$. As in the previous part, we assume a SNR of 20\,dB for the noisy UL channel vectors. Recall that an UDCT algorithm takes $N$ noisy UL  channel vectors as the input and produces an estimate of the DL covariance matrix as the output.  
We use the following distortion metrics:
\begin{enumerate}
	\item \textit{Normalized Frobenius-norm}: defined as 
	\[ E_{\text{NFD}} = \bE \left\{ \frac{\Vert \Sigdl -  {\Sigdl}^\star \Vert_\sfF}{\Vert \Sigdl\Vert_\sfF}\right\},  \]
	where $\|.\|_\sfF$ denotes the matrix Frobenius norm, where $\Sigdl$ and ${\Sigdl}^\star$ denote the true and the estimated DL covariance matrices, and where the expectation is taken over random  realizations of the  UL  channel vectors.

\item \textit{Power-loss}: Suppose that the eigenvalue decomposition of the true and estimated  DL covariance matrices are given as $\Sigdl= \Um \Dm \Um^\herm$ and $\Sigdl^\star= \widetilde{\Um} \widetilde{\Dm} \widetilde{\Um}^\herm$, respectively. We define the  efficiency metric of order $q\in [M]$  as
\begin{equation}\label{eq:efficiency_def}
\eta_q=\frac{\tr \left( \widetilde{\Um}_{q}^\herm \Sigdl\widetilde{\Um}_{q}\right)}{\tr \left(\Um_{q}^\herm\Sigdl\Um_{q}\right)}\in [0,1],
\end{equation}
where $\tr(.)$ denotes the trace operator, and 
where $\bfU_q$ and $\widetilde{\bfU}_q$ denotes $M \times q$ matrices consisting of the first $q$ columns of $\bfU$ and $\widetilde{\bfU}$ respectively.
This metric indicates which fraction of the power lying in the dominant $q$-dim subspace of the true DL covariance $\Sigdl$  is captured by the estimated subspace with the same dimension in $\Sigdl^\star$ spanned by $\widetilde{\Um}_{q}$ \cite{haghighatshoar2016massive,haghighatshoar2018low}. 
The closer the efficiency parameter $\eta_q$ is to one, the better the power in the dominant $q$-dim subspace of $\Sigdl$ is captured by the estimated covariance $\Sigdl^\star$. As a worst-case distortion metric that works independent of the dimension of the subspace, we consider 
\begin{align}
 E_{\text{PLE}} = 1- \bE \{ \min_{q \in [M]} \eta_q  \}
\end{align}
where the expected value is taken with respect to the random realization of the noisy UL  channel vectors.

\end{enumerate}

\begin{figure}[t]
	\centering
	\includegraphics[scale=1]{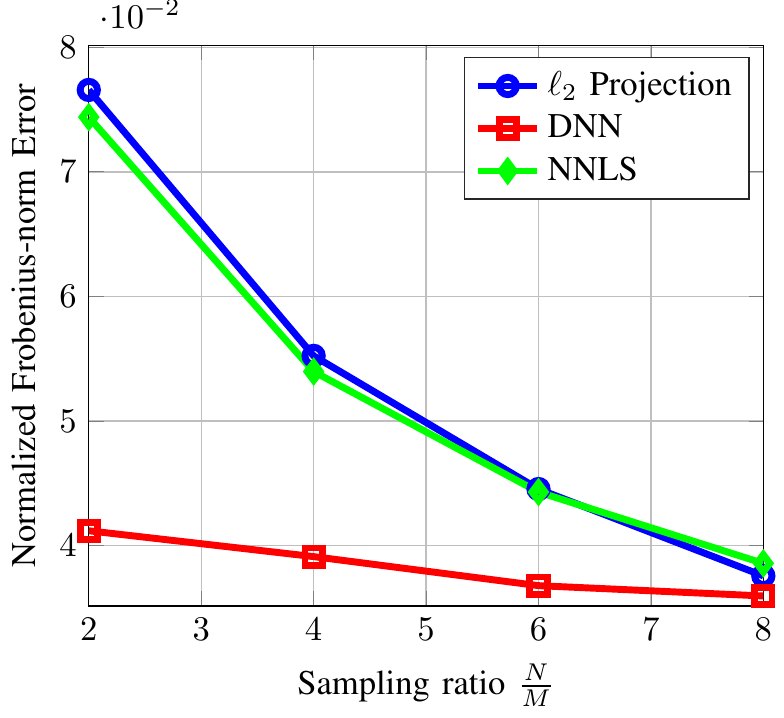}
	\caption{Comparison of the performance of UDCT for DNN, NNLS, and $\ell_2$ Projection method under Normalized Frobenius-norm distortion.}
	\label{fig:UDCT_fro}
\end{figure}

\begin{figure}[t]
	\centering
	\includegraphics[scale=1]{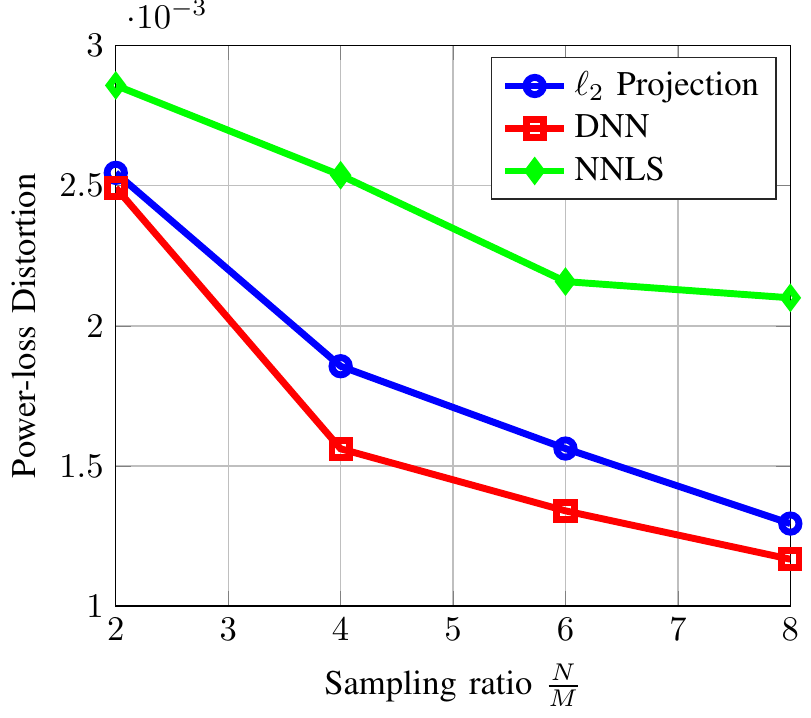}
	\caption{Comparison of the performance of UDCT for DNN, NNLS, and $\ell_2$ Projection method under Power-loss distortion.}
	\label{fig:UDCT_plos}
\end{figure}

Figs.\,\ref{fig:UDCT_fro} and \ref{fig:UDCT_plos} illustrate the simulation results. It is again seen that, as in the ASF estimation, our proposed method yields better performance in terms of UDCT under both distortion metrics, especially in the practically-relevant regime where the sampling ratio $\frac{N}{M}$ may be very small.

\subsection{Sparsifying Precoder vs. Statistical Beamforming}
In this paper, we designed  the sparsification precoding from the estimated DL covariance matrices $\Cdlk$, $k\in [K]$ (for simplicity, we dropped the label `dl'). In the presence of channel sparsification imposed by this precoder, BS still needs to probe the reduced-dim channel in the DL and request the users to feedback their channel measurements during the probing to the BS, from which the BS is able to extract the instantaneous channel state of the users after sparsification. 

As an alternative to sparsification precoding, one may use the statistical beamforming, which uses only the DL covariance knowledge to beamform to the users, and especially does not apply any DL probing and UL feedback to extract the instantaneous channel of the users \cite{wajid2009robust,zhang2015statistical}. Consequently, the statistical beamforming does not suffer from the fractional rate loss factor $1-\frac{\Tdl}{T}$ in \eqref{eq:rate_ub} caused because of devoting $\Tdl$ out of $T$ REs to DL channel probing.

We compare the performance of the sparsification precoder with the statistical beamforming method proposed in \cite{zhang2015statistical}. Let $\{\Sigmam_k\}_{k=1}^K$ denote the DL covariance matrices of the users, either exact or estimated via a UDCT method. In particular, in this section, we provide results for both the case in which the BS has access to exact DL covariances and the case where the DL covariances are estimated using DNN. Given $\{\Sigmam_k\}_{k=1}^K$, the beamforming vector for the user $k$ is given by 

\begin{equation}
\uv_k = \uv_{\max} \left \{ \big(N_0\mathbf{I}_M + \sum_{\ell \neq k} \Sigmam_\ell \big)^{-1} \Sigmam_k \right \},
\end{equation}
where $\uv_{\max}(.)$ denotes the normalized eigenvector corresponding to the maximum eigenvalue  of  a matrix, and where $N_0$ denotes the normalized noise power. The statistical beamforming matrix is then given by $\Vm= \left[ \uv_1,\ldots,\uv_K \right]$. We also assume uniform power allocation across the users, just as explained in Section \ref{sec:ch_sparsification}, and calculate the sum-rate according to \eqref{eq:rate_ub}.


\begin{figure}[t]
	\centering
	\includegraphics[scale=0.85]{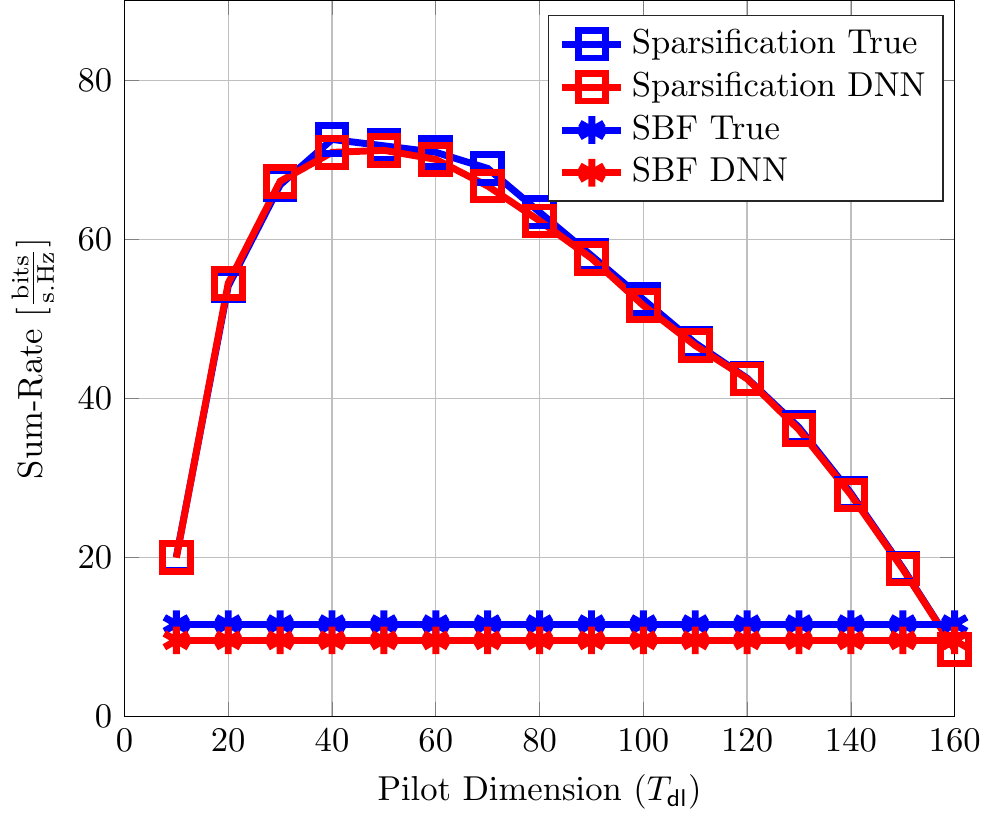}
	\caption{Sum-rate comparison of the proposed sparsification precoder with statistical beamforming for  $\SNR =20$ dB. We have $M=256$ BS antennas and $K=20$ served users. }
	\label{fig:rate_vs_Tdl}
\end{figure}

We set the simulation parameters as follows: the BS has  $M=256$ antennas and serves $K=20$ users, each having a randomly generated ASF as before. Fig.\,\ref{fig:rate_vs_Tdl} illustrates the sum-rate vs. pilot dimension results. The blue curves represent the results by assuming true DL covariances, and the red curves represent the results for estimated DL covariances via DNN. As we can see, the performance of the statistical beamforming (SBF) method is substantially inferior to the performance of the sparsification precoding  for almost all range of pilot dimensions. This implies that, it is always worthwhile to spend some of resource blocks to estimate instantaneous channels and design the beamformer upon them. This is true, even for small values of the pilot dimension, which implies that even a coarse estimation of the instantaneous channel using a few pilots yields a better performance  than the that of statistical beamforming.

\section{Conclusion}
In this paper, we presented a new approach to achieve a very competitive tradeoff between spatial multiplexing gain and probing-feedback overhead in FDD massive MIMO systems. Our approach has two main ingradients: (i)\,an efficient regularization technique  based on Deep Neural Networks (DNN) for structured ASF estimation and UL-DL covariance transformation, (ii)\,a novel ``\textit{sparsifying precoding}'' technique that uses the estimated DL covariance matrices from  (i) and shapes the channel sparsity depending on the pilot dimension such that the ``\textit{effective}'' channel vectors  after sparsification can be estimated at the BS with a low mean-square error. 
In particular, the proposed sparsifying precoder is not at the mercy of the channel sparsity induced by the propagation environment, which is a bottleneck for Compressed Sensing based channel estimation methods.
	We compared our proposed DNN-based method in (i) with other methods  in the literature via numerical simulations  and showed that it yields a very competitive performance.  We also compared our sparsifying precoder in (ii)  with the state-of-the-art statistical beamforming methods under the assumption that  those methods also have access to the covariance knowledge in the DL and showed that our method yields higher spectral efficiency since it uses in  addition the instantaneous channel information after sparsification.

	\balance
	
	{\small
		\bibliographystyle{IEEEtran}
		\bibliography{references2}
	}
	
\end{document}

%% file: symbols.tex
\usepackage{mathbbol}

\def\mindex#1{\index{#1}}

\def\sq{\hbox{\rlap{$\sqcap$}$\sqcup$}}
\def\qed{\ifmmode\sq\else{\unskip\nobreak\hfil
\penalty50\hskip1em\null\nobreak\hfil\sq
\parfillskip=0pt\finalhyphendemerits=0\endgraf}\fi\medskip}

\long\def\defbox#1{\framebox[.9\hsize][c]{\parbox{.85\hsize}{%
\parindent=0pt
\baselineskip=12pt plus .1pt      
\parskip=6pt plus 1.5pt minus 1pt 
 #1}}}

\long\def\beginbox#1\endbox{\subsection*{}%
\hbox{\hspace{.05\hsize}\defbox{\medskip#1\bigskip}}%
\subsection*{}}

\def\endbox{}

\def\tr{\mathsf{tr}}

\newsavebox{\junk}
\savebox{\junk}[1.6mm]{\hbox{$|\!|\!|$}}

\def\bC{{\mathbb C}}

\def\bE{{\mathbb E}}

\def\bG{{\mathbb G}}

\def\bR{{\mathbb R}}

\def\bfH{{\bf H}}

\def\bfU{{\bf U}}

\def\bfX{{\bf X}}

\def\bfa{{\bf a}}

\def\bfd{{\bf d}}

\def\bfh{{\bf h}}

\def\bfx{{\bf x}}
\def\bfy{{\bf y}}
\def\bfz{{\bf z}}

\def\scrC{{\mathscr{C}}}

\def\sfF{{\sf F}}

\def\sfH{{\sf H}}

\def\bfmath#1{{\mathchoice{\mbox{\boldmath$#1$}}%
{\mbox{\boldmath$#1$}}%
{\mbox{\boldmath$\scriptstyle#1$}}%
{\mbox{\boldmath$\scriptscriptstyle#1$}}}}

\def\bfmY{\bfmath{Y}}

\def\bfmhhaY{\bfmath{\hhaY}} 
\def\bfmhhaY{\hbox to 0pt{$\widehat{\bfmY}$\hss}\widehat{\phantom{\raise 1.25pt\hbox{$\bfmY$}}}}

\def\til={{\widetilde =}}

\def\clA{{\cal A}}

\def\clC{{\cal C}}

\def\clG{{\cal G}}
\def\clH{{\cal H}}

\def\clN{{\cal N}}

\def\clX{{\cal X}}

 \def\FRAC#1#2#3{\genfrac{}{}{}{#1}{#2}{#3}}

\def\ddtp{{\mathchoice{\FRAC{1}{d^{\hbox to 2pt{\rm\tiny +\hss}}}{dt}}%
{\FRAC{1}{d^{\hbox to 2pt{\rm\tiny +\hss}}}{dt}}%
{\FRAC{3}{d^{\hbox to 2pt{\rm\tiny +\hss}}}{dt}}%
{\FRAC{3}{d^{\hbox to 2pt{\rm\tiny +\hss}}}{dt}}}}

\def\average#1,#2,{{1\over #2} \sum_{#1}^{#2}}

\def\eye(#1){{\bf(#1)}\quad}

\def\eq#1/{(\ref{e:#1})}

\newcommand{\beqn}[1]{\notes{#1}%
\begin{eqnarray} \elabel{#1}}

\newcommand{\eeqn}{\end{eqnarray} }

\newcommand{\beq}[1]{\notes{#1}%
\begin{equation}\elabel{#1}}

\newcommand{\eeq}{\end{equation}}

\def\bdes{\begin{description}}
\def\edes{\end{description}}

\newcounter{rmnum}

\newcounter{anum}

{\end{list}}

\def\ass(#1:#2){(#1\ref{#1:#2})}

\def\ritem#1{
\item[{\sf \ass(\current_model:#1)}]
}

\newenvironment{recall-ass}[1]{%
\begin{description}
\def\current_model{#1}}{
\end{description}
}



%% file: macros.tex
\long\def\comment#1{}

\newcommand{\fv}{{\bf f}}

\newcommand{\hv}{{\bf h}}

\newcommand{\uv}{{\bf u}}

\newcommand{\yv}{{\bf y}}

\newcommand{\Am}{{\bf A}}
\newcommand{\Bm}{{\bf B}}

\newcommand{\Dm}{{\bf D}}

\newcommand{\Fm}{{\bf F}}
\newcommand{\Gm}{{\bf G}}
\newcommand{\Hm}{{\bf H}}
\newcommand{\Id}{{\bf I}}

\newcommand{\Lm}{{\bf L}}

\newcommand{\Pm}{{\bf P}}

\newcommand{\Um}{{\bf U}}
\newcommand{\Wm}{{\bf W}}
\newcommand{\Vm}{{\bf V}}

\newcommand{\Ac}{{\cal A}}
\newcommand{\Bc}{{\cal B}}

\newcommand{\Ec}{{\cal E}}

\newcommand{\Kc}{{\cal K}}
\newcommand{\Lc}{{\cal L}}
\newcommand{\Mc}{{\cal M}}
\newcommand{\Nc}{{\cal N}}

\newcommand{\Pc}{{\cal P}}

\newcommand{\Sc}{{\cal S}}

\newcommand{\lambdav}{\hbox{\boldmath$\lambda$}}

\newcommand{\Lambdam}{\hbox{\boldmath$\Lambda$}}

\newcommand{\Sigmam}{\hbox{\boldmath$\Sigma$}}

\newcommand{\Psim}{\hbox{\boldmath$\Psi$}}

\newcommand{\trace}{{\hbox{tr}}}

\newcommand{\SNR}{{\sf SNR}}

\newcommand{\transp}{{\sf T}}

